%% file: manuscript.tex
\documentclass[prb,reprint,superscriptaddress,longbibliography]{revtex4-1}
\usepackage{graphicx}
\usepackage[caption=false]{subfig}
\usepackage{bm}
\usepackage{amsmath}
\usepackage{amssymb}
\usepackage{mathtools}
\usepackage{siunitx}
\usepackage{multirow}
\usepackage{array}
\usepackage{xcolor}
\usepackage{longtable}
\usepackage[version=4]{mhchem}
\usepackage{hyperref}
\hypersetup{colorlinks=true,
            linkcolor=red,
            citecolor=blue,
            urlcolor=orange}
 
\newcolumntype{K}[1]{>{\centering\arraybackslash}p{#1}}
\setcitestyle{numbers,round}

\AtBeginDocument{\usepackage{booktabs}}

\begin{document}

\title{Exploring the high-pressure materials genome}

\author{Maximilian Amsler}
\email{amsler.max@gmail.com}
\altaffiliation{Present address: Laboratory of Atomic and Solid State Physics,
Cornell University, Ithaca, New York 14853, USA}
\affiliation{Department of Materials Science and Engineering, Northwestern
University, Evanston, Illinois 60208, USA}

\author{Vinay I. Hegde}
\email{hegde@u.northwestern.edu}
\affiliation{Department of Materials Science and Engineering, Northwestern
University, Evanston, Illinois 60208, USA}

\author{Steven D. Jacobsen}
\email{s-jacobsen@northwestern.edu}
\affiliation{Department of Earth and Planetary Sciences, Northwestern University, Evanston, Illinois 60208, USA}

\author{Chris Wolverton}
\email{c-woverton@northwestern.edu}
\affiliation{Department of Materials Science and Engineering, Northwestern
University, Evanston, Illinois 60208, USA}

\date{\today}

\begin{abstract}
\input{abstract}
\end{abstract}

\maketitle


\section{Introduction}\label{sec:introduction}
\input{introduction}



\input{method}
\section{Results and Discussion}\label{sec:results}
\input{results}
\section{Conclusions}\label{sec:conclusions}
\input{conclusions}

\section{Methods}\label{sec:methods2}
\input{method2.tex}


\section{Acknowledgments}\label{sec:ack}
\input{acknowledgements}

\bibliographystyle{apsrev4-1}
%

\end{document}

%% file: abstract.tex
A thorough \textit{in situ} characterization of materials at extreme conditions is challenging, and computational tools such as crystal structural search methods in combination with \textit{ab initio} calculations are widely used to guide experiments by predicting the composition, structure, and properties of high-pressure compounds. However, such techniques are usually computationally expensive and not suitable for large-scale combinatorial exploration. On the other hand, data-driven computational approaches using large materials databases are useful for the analysis of energetics and stability of hundreds of thousands of compounds, but their utility for materials discovery is largely limited to idealized conditions of zero temperature and pressure. Here, we present a novel framework combining the two computational approaches, using a simple linear approximation to the enthalpy of a compound in conjunction with ambient-conditions data currently available in high-throughput databases of calculated materials properties. We demonstrate its utility by explaining the occurrence of phases in nature that are not ground states at ambient conditions and estimating the pressures at which such ambient-metastable phases become thermodynamically accessible, as well as guiding the exploration of ambient-immiscible binary systems via sophisticated structural search methods to discover new stable high-pressure phases.

%% file: introduction.tex
The laws of thermodynamics dictate that only compounds corresponding to global
minima of the Gibbs free energy for a given set of external conditions are
viable ground states with infinite lifetimes~\cite{gibbs_equilibrium_1878}. For
such materials, there always exists a synthetis route that follows an overall
exothermic chemical reaction pathway, and all systems at finite temperature will
ultimately attain a Boltzmann distribution with a high occupation of the ground
state in thermodynamic equilibrium. In practice, however, materials in many
industrially relevant applications are metastable, i.e., they have higher
energies than the equilibrium ground states. Such metastable phases, or
polymorphs, correspond to local minima on the energy landscape and are
surrounded by sufficiently high barriers to render them kinetically persistent
on a finite time scale~\cite{rosenstein_aspects_1969, brog_polymorphism_2013}.

Synthesizing metastable materials essentially requires finding, in some manner,
a path in configurational space such that precursors undergo chemical reactions
along a downhill trajectory with sufficiently low activation barriers, until the
desired product is formed and quenched~\cite{gopalakrishnan_chimie_1995,
stein_turning_1993}. A plethora of thermodynamic parameters can be tuned to
design such a pathway, including temperature, pressure, electromagnetic fields,
compositional variations, choosing specific precursor materials, etc. A special
case of this design procedure is to choose a set of thermodynamic parameters
such that the desired phase becomes the thermodynamic ground state at the chosen
conditions, where it forms at equilibrium, and can be recovered as a metastable
phase at ambient conditions if all transition barriers leading away from it are
sufficiently high~\cite{sun_thermodynamic_2016}.

This problem of identifying the ground states for a given set of external
conditions is commonly tackled in the computational materials discovery
community through global optimization of a target fitness function, using
advanced crystal structure prediction (CSP) methods~\cite{oganov_modern_2010}.
Ideally, this fitness function corresponds to the Gibbs free energy, but it is
often approximated by the potential energy (at zero pressure, temperature) or
the enthalpy (at zero temperature) or some other biased energy landscape,
and is sampled in an unconstrained manner in the configurational space. Many
novel materials and their structures have been resolved using CSP at high
pressures~\cite{pickard_ab_2011, zhang_unexpected_2013, walsh_discovery_2016,
zhang_materials_2017}, using chemical pressure and thermal
degassing~\cite{amsler_low-density_2015, rasoulkhani_energy_2017}, as
2-dimensional materials~\cite{wang_effective_2012, zhou_semimetallic_2014,
eivari_two-dimensional_2017}, or at surfaces and
interfaces~\cite{chua_genetic_2010, schusteritsch_predicting_2014,
lu_self-assembled_2014, zhao_interface_2014, fisicaro_surface_2017}. However,
CSP approaches are computationally demanding and their applications are
therefore often limited to small subsets of chemical spaces.

On the other hand, data-driven approaches using large materials databases in
conjunction with high-throughput (HT) density functional theory (DFT)
calculations have become increasingly popular in materials
science~\cite{saal_materials_2013, kirklin_oqmd_2015, Jain2013,
curtarolo_aflowlib.org_2012, curtarolo_high-throughput_2013}. Such HT databases
usually contain DFT-calculated properties such as formation energy, equilibrium
volume, and relaxed atomic coordinates for experimentally reported phases
available in repositories such as the Inorganic Crystal Structure Database
(ICSD)~\cite{icsd}. These datasets are sometimes augmented with hypothetical
compounds constructed by decorating common structural prototypes with elements
in the periodic table. Subsequent phase stability analysis is often performed to
identify stable phases in every chemical space. Although approaches using such
HT-DFT databases are useful for efficient large-scale analysis of energetics
across a wide range of chemistries, they lack the power to predict novel
materials with unknown crystal structures, and phases beyond ambient conditions
since all such databases currently contain only materials properties calculated
at zero temperature and zero pressure.

In this work, we effectively combine big-data in HT-DFT databases with CSP
methods to predict and discover novel materials stable at non-ambient
conditions. Using the ``implicitly available'' high-pressure information in a
HT-DFT database, the Open Quantum Materials Database (OQMD), together with a
simple approximation to the formation enthalpy of a compound, we study the
effect of pressure on the thermodynamic scale of stability/metastability of
inorganic compounds. Our model correctly predicts most (75--80\%) experimentally
reported high-pressure elemental and binary phases to become thermodynamically
stable at non-ambient pressures. In fact, our statistical analysis of the data
in the OQMD shows a large fraction of ambient-metastable compounds to be
thermodynamic ground states at non-zero pressures. From an experimental point of view, vast unexplored
pressure-composition space is becoming widely accessible through diamond-anvil cell
techniques~\cite{shen_high-pressure_2017}, and improving predictive methods for high-pressure phases is, e.g., relevant
to geophysical studies of planetary interiors where there can be numerous polymorphs energetically in close proximity,
even in relatively simple compositional systems~\cite{tsuchiya_first_2013}. Here, we use our model to sample
\textit{all} binary intermetallic chemical spaces with no experimentally reported compound in the OQMD ($\sim$1780
chemical spaces) and predict nearly 3800 new compounds to be stable at some finite pressure. Finally, we demonstrate the
power of our predictive framework in guiding sophisticated CSP methods by explicitly exploring ten binary-immiscible
systems, and discover that our model correctly predicts phase spaces containing novel high-pressure materials, which
could be potentially recovered to ambient conditions as metastable compounds.

%% file: method.tex

Let us introduce a model to approximate the enthalpy of a phase to efficiently evaluate the phase stability of hundreds of thousands of compounds in a large chemical space at arbitrary pressure.  
\subsection{Linear approximation to enthalpy}\label{ssec:linear_approximation_to_enthalpy}
At zero temperature and pressure $p$, the Gibbs free energy for a given phase
reduces to the enthalpy $H=E+pV$, where $E$ is the internal energy and $V$ is
the volume of the phase. Expanding $H$ as a function of $p$ around the
equilibrium pressure $p_0$ yields~\cite{pickard2011}
\begin{align}\label{eqn:enthalpy_expansion}
H(p) & = H(p_0) + \Delta p\,H'(p_0) + \frac{(\Delta p)^2}{2}H''(p_0) +\, \cdots
\nonumber \\
     & = H(p_0) + \Delta p\,V(p_0) + \frac{(\Delta p)^2}{2}\frac{V(p_0)}{B(p_0)}
     +\, \cdots
\end{align}
where $\Delta p = (p - p_0)$, and $B=\frac{1}{\beta}$ is the bulk modulus of the
phase, where $\beta=-\frac{1}{V} \frac{\partial V}{\partial p}$ is its
compressibility. If we neglect all terms higher than second order and consider
all phases to be incompressible (i.e., $B(p_0) \to \infty$), for equilibrium
pressure $p_0 = 0$, we can approximate the enthalpy of a phase simply as
\begin{equation}\label{eqn:enthalpy_approximation}
 H(p) = E_0 + \Delta p\,V(p_0)
\end{equation}
where $E_0$ is the internal energy at the equilibrium volume $V_0$.
Conveniently, both $E_0$ and $V_0$ are quantities that are readily available for
hundreds of thousands of phases in most HT-DFT materials databases such as the
OQMD~\cite{saal_materials_2013, kirklin_oqmd_2015}, Materials
Project~\cite{Jain2013}, and AFLOWlib~\cite{curtarolo_aflowlib.org_2012}. 


\begin{figure*}[!htbp]
\centering
\includegraphics[width=1\textwidth]{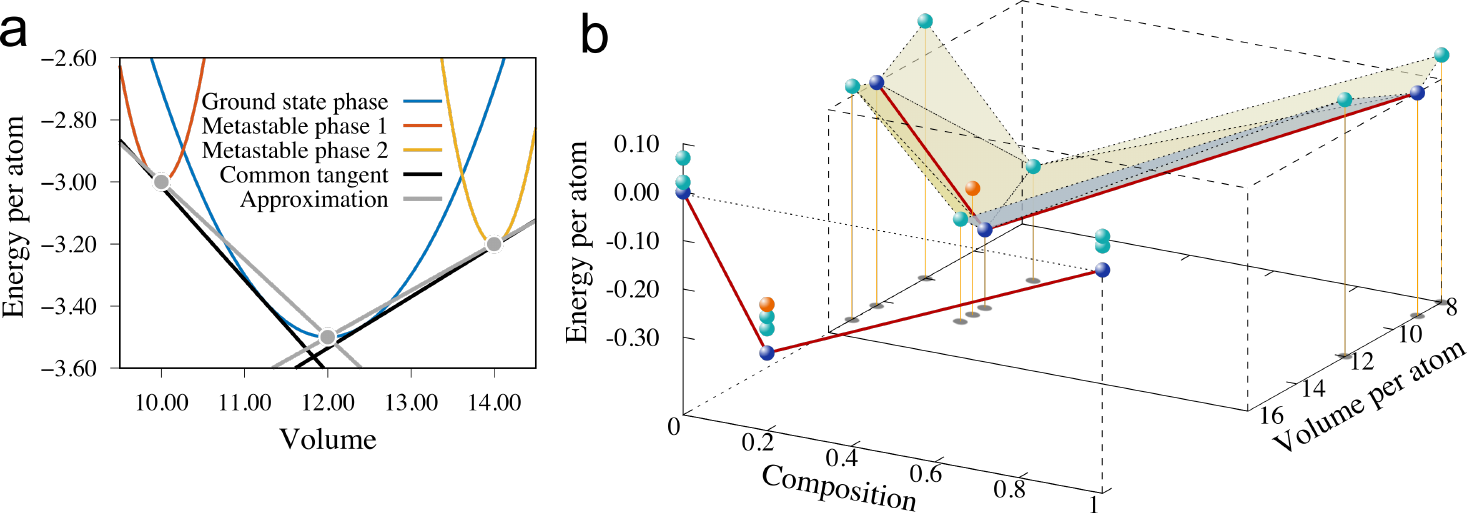}
\caption{(a) A schematic energy-volume ($V$--$E$) diagram with three phases and
their EOSs, each represented by a parabola. The negative slope of the common
tangent to two adjacent EOS (solid black line) represents the pressure at which
the two phases are in equilibrium. The LAE approximates the common tangent with a line
connecting the ambient-condition equilibrium volumes/energies of two adjacent
phases (solid grey line connecting filled grey circles). (b) A schematic
$N$--$V$--$E$ convex hull for a model binary system. Individual phases are
represented by spheres, and convex hull boundaries are indicated with solid red
and dotted black lines. On the left is the conventional zero-pressure $N$--$E$
hull, a projection of the extended $N$--$V$--$E$ convex hull on the right. Phases
that are thermodynamically stable at zero pressure lie on the $N$--$E$ convex
hull (blue spheres). Metastable phases that are stable at some non-ambient
pressure lie \textit{above} the $N$--$E$ hull but \textit{on} the $N$--$V$--$E$
convex hull (teal spheres). A phase that is truly unstable at any pressure lies
above the $N$--$V$--$E$ hull (orange sphere).}\label{fig:hulls}
\end{figure*}

The above linear approximation to enthalpy (henceforth referred to as ``LAE'')
is illustrated in an energy-volume diagram in Fig.~\ref{fig:hulls}a, where the
ground state and two metastable states are each represented by their respective
equation of states (EOS) $E(V)$, i.e., their energy as a function of volume,
approximated by parabola. The negative slopes of the common tangents connecting
the EOS of neighboring phases represent the pressure at which both phases are in
equilibrium (``transition pressures'', black solid lines). With our
approximation of the bulk moduli $B(p_0) \to \infty$, the EOS curve of each
phase would have infinitely large curvature, reducing the parabola to a vertical
line originating at the corresponding equilibrium volumes $V_0$ and energies
$E_0$. Essentially, all information of each phase is then contained in a single
point at ($V_0$, $E_0$), represented by filled points.

Although the LAE is rather crude, it is reasonably accurate up to pressures in
the range of tens, or even hundreds of GPa. As we will show in the rest of this
work, the LAE can be used as a powerful tool to enable quick analyses of phase
stability of a large number of materials at non-ambient pressures. Note that we
will hereafter use the terms ``zero pressure'' and ``ambient pressure''
interchangeably, since the $pV$ contribution to the free energy at atmospheric
pressure is insignificant for most inorganic compounds. E.g., at one atmosphere,
which corresponds to roughly 1 bar, the energy contribution of $pV$ in diamond silicon
with a volume of $\sim$20~\AA$^3$/atom is merely 0.012~meV/atom, far smaller
than the error bars encountered in DFT calculations.


\subsection{Thermodynamic stability: the convex hull}\label{ssec:thermodynamic_stability_convex_hull}

The thermodynamic stability of a phase at zero temperature can be determined
by the construction of the so-called convex hull of all phases in the chemical
space. At zero pressure, the convex hull is constructed from the composition and
formation energy (composition-energy hull, or simply ``$N$--$E$ convex hull'')
of all the phases. By definition, a phase \textit{on} the convex hull has a
formation energy lower than that of any other phase (or linear combination of
phases) at that composition, and is therefore thermodynamically stable. At
non-ambient pressures, thermodynamic stability is determined by a convex hull
which also takes into account the energy as a function of volume of all phases,
given by their respective EOS $E(V)$. The LAE introduced in
Section~\ref{ssec:linear_approximation_to_enthalpy} allows us to simplify the
construction of the convex hull by taking into account the ambient volume of
each phase, in addition to their composition and formation energy
(compositon-volume-energy hull, or simply ``$N$--$V$--$E$ convex hull''). A
phase on the extended $N$--$V$--$E$ hull has a formation energy lower than any
other phase or combination of phases at that composition \textit{and} volume,
and is therefore thermodynamically stable at some pressure. Further, a tie line
on the convex hull represents a two-phase equilibrium, a triangular facet
represents a three-phase equilibrium, and so on---a facet with $n$ vertices
represents an $n$-phase equilibrium.

A schematic $N$--$V$--$E$ convex hull is shown in Fig.~\ref{fig:hulls}b. A
projection of the extended $N$--$V$--$E$ convex hull taking into account only
the energy and volumes leads to the $N$--$E$ hull (indicated by solid red
lines). Phases that lie \textit{above} the $N$--$E$ hull, but \textit{on} the
$N$--$V$--$E$ hull, are metastable at zero pressure but thermodynamically stable
at some finite pressure. For example, in Fig.~\ref{fig:hulls}b, only two
elemental phases and one binary compound (blue spheres) lie on the $N$--$E$ hull
(solid red lines), i.e., are thermodynamically stable at zero pressure, and all
other phases are metastable. However, all elemental phases and all binary
compounds except one lie on the extended $N$--$V$--$E$ hull (teal spheres
connected by dotted black lines), i.e., are thermodynamically stable at some
non-ambient pressure. Only one phase shown (orange sphere at composition 0.2) is
truly unstable at all pressures.

\subsection{Pressure range of stability}\label{ssec:pressure_range_of_stability}

For a system in thermodynamic equilibrium at zero temperature, $dE = - p\,dV +
\sum_i \mu_i\,dN_i$, where $dE$, $dV$ are infinitesimal changes in internal
energy $E$, volume $V$ of the system, respectively, and $dN_i$ is the infinitesimal
change in the composition $N_i$ of species $i$. The
equilibrium pressure is thus given by $p = -\left(\frac{\partial E}{\partial
V}\right)_{N_i}$, i.e., the derivative of energy with respect to volume at
constant composition.  Hence, the pressure range of stability of a phase
$\mathcal{P}$ with ambient equilibrium volume and energy of $V_0$ and $E_0$,
respectively, is governed by the phase equilibria at volumes $(V_0 + dV)$ and
$(V_0 - dV)$~\footnote{This procedure is analogous to calculating the range
of thermodynamic stability of a compound with respect to the chemical potential
of a given species. Since the chemical potential of species $i$, at $p$, $T=0$,
is given by $\mu_i = \left(\frac{\partial E}{\partial N_i}\right)_{N_{j \ne
i}}$, the window of stability [$\mu_i^-$, $\mu_i^+$] of the phase can be
calculated using $\mu_i^{\pm} = \frac{E_0 - E(N_i \mp dN_i)}{dN_i}$, i.e., as a
perturbation of energy with respect to the composition of the species $i$.}. In
 other words, the window of pressures [$p_-$, $p_+$] where $\mathcal{P}$ is
stable is given by
\begin{equation}\label{eqn:pressure_window}
p_{\pm} = - \frac{E_0 - E(V_0 \mp dV)}{dV}
\end{equation}
$E(V_0 \pm dV)$ can be calculated by minimizing the free energy of the system at
the target composition \textit{and} volume. Grand canonical linear programming
(GCLP)~\cite{Akbarzadeh2007} techniques using efficient linear solvers are
routinely employed to calculate phase stabilities and equilibrium reaction
pathways at 0~K and 0~GPa~\cite{Mason2011, Aidhy2011, Kirklin2013, Emery2016,
Ma2017}. In this work, in addition to the average composition of the system
being constrained to that of $\mathcal{P}$, the volume is constrained to $V_0
\pm dV$ during energy minimization. Thus, a pressure range of stability can be
calculated for every phase that lies on the extended $N$--$V$--$E$ convex hull.

As discussed in Sec.~\ref{ssec:linear_approximation_to_enthalpy}, the negative
slope of the common tangent to the EOS of two phases is the pressure at which
the respective phases coexist, or in other words, one phase transforms into the
other under the effect of pressure. In the LAE, the common tangent is reduced to
a line connecting the local minima of the two phases (solid gray line connecting
filled grey circles in Fig.~\ref{fig:hulls}a). The LAE introduces errors compared to the real transition pressure, which depend on the overall features of the energy landscape. If we assume that all phases are \textit{compressible} with identical, \textit{finite} bulk moduli, the LAE will consistently lead to an \textit{underestimation} of the magnitude of the transition pressures. In practice, however, high-pressure phases often exhibit shorter, stronger bonds that lead to \textit{higher} bulk moduli. Hence, the LAE would lead to a better agreement with the real transition pressures for phases stable at very high pressures. On the other hand, if the bulk moduli significantly \textit{decreased} with pressure, the LAE would lead to an  \textit{overestimation} of the magnitude of the transition pressures. We also note that transition pressures, based on the above definition, can be positive or negative (e.g., the
common tangents connecting the ground state with metastable phases 1 and 2,
respectively in Fig.~\ref{fig:hulls}a). A negative pressure can be physically
interpreted as a tensile stress, leading to the expansion of a phase toward
volumes exceeding its ambient ground state equilibrium volume.

%% file: results.tex

\subsection{Model validation}\label{ssec:model_validation}

We first evaluate the accuracy of the linear approximation to enthalpy by
investigating two elements and three binary systems in detail.

\subsubsection{Elemental solids}\label{sssec:elemental_solids}
\input{validation_elements}


\subsubsection{Binary intermetallics}\label{sssec:binary_intermetallics}
\input{validation_binaries}


\subsection{Large-scale analysis of phase stability at high pressure}\label{ssec:statistical_analysis}

\subsubsection{Elements and binary compounds}\label{sssec:statistics_elements}
\input{statistics_elements}


\subsubsection{All experimentally reported compounds}\label{sssec:statistics_all}
\input{statistics_all}

\subsection{Discovery of new high-pressure compounds}\label{ssec:prediction}
\input{binary_immiscible}

%% file: validation_elements.tex
We choose two elements whose high-pressure phase diagrams are among the most
complex as well as the most well-studied: silicon and bismuth. Both elements
have intricate energy landscapes with several high-pressure allotropes.\\

\noindent
\textit{a. Silicon}\\

\noindent The phase diagram of silicon has been well explored experimentally,
partially due to its importance in the semiconductor industry. The ambient
ground state is Si-I, which crystallizes in a cubic diamond
structure~\cite{dutta_lattice_1962}. It transforms around 11~GPa to the Si-II
phase, which has a $\beta$-Sn structure~\cite{jamieson_crystal_1963}. This is
followed by a transformation to Si-XI with $Imma$
symmetry~\cite{mcmahon_new_1993} at 13~GPa. Above 16~GPa, Si-V forms in the
simple hexagonal structure~\cite{hu_phases_1984}, and at 38~GPa, Si-VI forms in
an orthorhombic $Cmcm$ structure~\cite{hanfland_crystal_1999}. The hexagonal
close-packed Si-VII forms above 42~GPa~\cite{olijnyk_structural_1984}, and
finally, cubic close-packed Si-X forms at pressures above
78~GPa~\cite{duclos_hcp_1987}.

\begin{figure}[!htbp]
\centering
\includegraphics[width=0.9\columnwidth,angle=0]{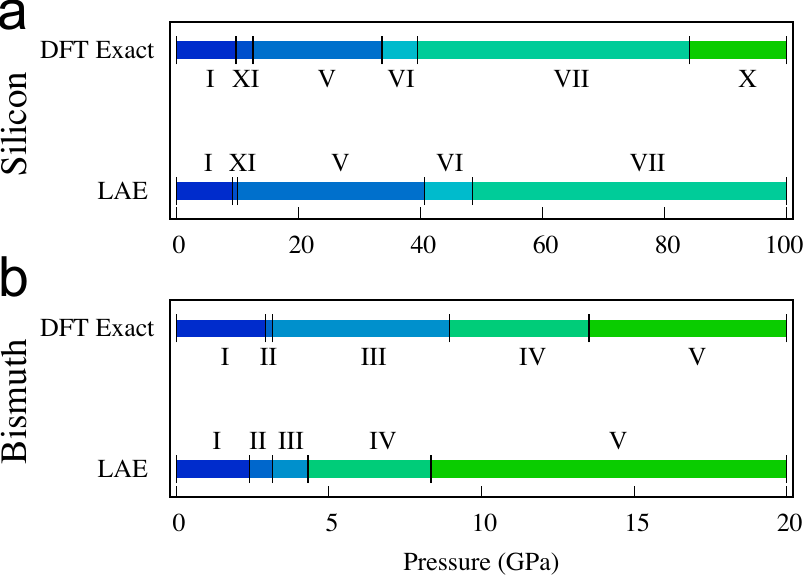}
\caption{The pressure range of stability of high-pressure phases of elemental
(a) silicon and (b) bismuth, respectively. Explicitly computed transition
pressures using DFT-calculated formation enthalpies are labeled ``DFT Exact''
(top bar in each panel), and those based on the LAE are denoted with  ``LAE''
(bottom bar in each panel). The crystal structures of the silicon and bismuth
allotropes were taken from the Refs.~\onlinecite{dutta_lattice_1962,
jamieson_crystal_1963, mcmahon_new_1993, hu_phases_1984, hanfland_crystal_1999,
olijnyk_structural_1984, duclos_hcp_1987} and~\onlinecite{brugger_crystal_1967,
akselrud_refinement_2014,chen_crystal_1994, mcmohan_incommensurate_2000,
haussermann_comparative_2002,chen_structural_1997,aoki_stability_1982},
respectively.}\label{fig:elements}
\end{figure}

We first compute the pressure range of stability of the various silicon
allotropes using DFT calculations (see the top panel labeled ``Exact'' in
Fig.~\ref{fig:elements}(a)). For each phase, we calculate the enthalpy
explicitly at various pressures at intervals of 2~GPa and 10~GPa in the range of
0--20~GPa and 20--100~GPa, respectively. The transition pressures are then
computed by minimizing the interpolated formation enthalpies as a function of
pressure. The experimentally reported sequence of formation and transition
pressures of high-pressure Si allotropes are well reproduced, with the exception
of Si-II, which is effectively degenerate in enthalpy to Si-XI. The discrepancy
between experiment and theory for the transition from Si-I to Si-II has been
well studied~\cite{sun_accurate_2016, hennig_phase_2010}, and is attributed to
the errors associated with the PBE approximation to the DFT exchange correlation
potential.
 

We then calculate the pressure range of stability of all the allotropes using
only the respective equilibrium energies and volumes at 0~GPa, extrapolated
linearly as described in
Sections~\ref{ssec:linear_approximation_to_enthalpy}--\ref{ssec:pressure_range_of_stability}
(see the bottom panel labeled ``LAE'' in Fig.~\ref{fig:elements}a). The
agreement between the ``DFT Exact'' and ``LAE'' phase diagrams is remarkable:
(a) the sequence of the phases is correctly reproduced, with the only exception
of Si-X, which the linear approximation model predicts to be unstable even at
100~GPa, and (b) the overall errors in the transition pressures predicted by the
approximate model are within around 10\% of those calculated explicitly.\\

\noindent
\textit{b. Bismuth}\\

\noindent
At ambient condition, bismuth crystallizes in a rhombohedral Bi-I phase with
space group $R\bar{3}m$. It transforms at a pressure of around 2.55~GPa to Bi-II
with a $C2/m$ structure~\cite{brugger_crystal_1967, akselrud_refinement_2014}
and a very narrow range of stability at low temperatures. Upon increasing the
pressure, Bi-III forms in a complicated, incommensurate host-guest structure
with $P4/ncc$ symmetry~\cite{chen_crystal_1994, mcmohan_incommensurate_2000,
haussermann_comparative_2002}. A Bi-IV phase with space group $P21/n$ has been
reported between 2.4~GPa and 5.3~GPa at temperatures above around
450~K~\cite{chen_structural_1997}. Finally, the Bi-V bcc phase is observed at
pressures above 7.7~GPa~\cite{aoki_stability_1982}.

Similar to the case of silicon, we first compute the pressure range of stability
of the various bismuth allotropes using enthalpies calculated explicitly at
various pressures at intervals of 1~GPa in the range of 0--20~GPa (see the top
panel labeled ``DFT Exact'' in Fig.~\ref{fig:elements}b). Although the
experimentally reported sequence of allotropes formed is well reproduced, the
transition pressures between Bi-III/Bi-IV and Bi-IV/Bi-V are severely
overestimated. This behavior has been reported previously by
H{\"a}ussermann~et~al.~\cite{haussermann_comparative_2002}, and corroborated in
our recent work on Cu--Bi intermetallics~\cite{clarke_discovery_2016,
clarke_creating_2017}.


The pressure range of stability of all allotropes calculated using the LAE
reproduces the correct sequence of formation (see the bottom panel labeled
``LAE'' in Fig.~\ref{fig:elements}b). However, the agreement between the
transition pressures predicted by the approximate model and those calculated
explicitly are worse than that for silicon allotropes. We attribute these larger
errors to the strong changes in the chemical bonds between the different bismuth
phases, especially since ambient Bi-I has a layered structure, in contrast to
the high-pressure phases. Hence, our approximation of equal, infinitely large
bulk moduli for every phase is perhaps less reasonable for elemental phases of
bismuth.


%% file: validation_binaries.tex
When compared to pure elements, the high-pressure phase space of
binary/higher-order chemical systems have been experimentally relatively
unexplored. Including composition and pressure as additional degrees of freedom
significantly increases the complexity of the phase space. In this section, we
focus on a unique subset of chemistries: intermetallic systems of elements that
are not miscible at ambient conditions but form compounds under pressure. Many
of these so-called ambient-immiscible systems involve bismuth in combination
with other elements. Recently, we investigated three such systems in detail,
namely Fe--Bi~\cite{amsler_prediction_2017}, Cu--Bi~\cite{clarke_discovery_2016,
clarke_creating_2017}, and Ni--Bi~\cite{powderly_high-pressure_2017}, by
performing extensive global structure searches. Here, we use these three systems
to further evaluate the performance of the linear approximation to enthalpy.\\

\begin{figure*}[!htbp]
\centering
\includegraphics[width=\textwidth,angle=0]{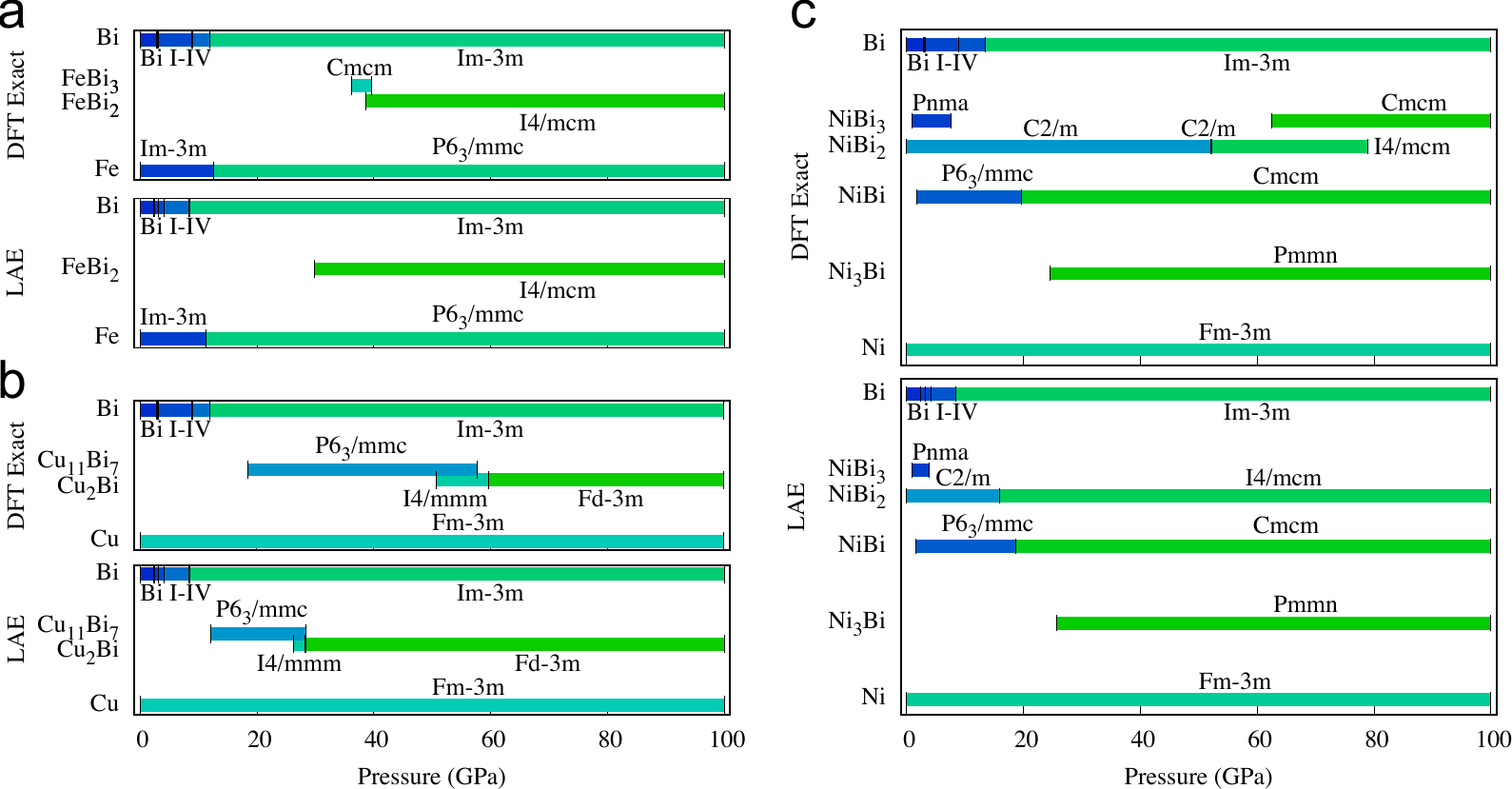}
\caption{Comparison between the explicitly computed phase diagrams with the ones
derived from the LAE model for binary systems. The panels (a), (b) and (c)
correspond to the Fe--Bi, Cu--Bi and Ni--Bi systems, respectively.  Explicitly
calculated transition pressures using DFT are denoted with ``DFT Exact'' (top bar),
and results based on the LAE are denoted with  ``LAE'' (bottom
bar).}\label{fig:binaries}
\end{figure*}

\noindent
\textit{a. Fe--Bi}\\
\noindent
Using the minima hopping crystal structure prediction method (MHM), we recently predicted a high-pressure \ce{FeBi2}
phase  with \textit{I4/mcm} symmetry at pressures above
36~GPa~\cite{amsler_prediction_2017}, which was  experimentally confirmed through evidences 
 found in the \textit{in-situ} X-ray diffraction pattern at above
30~GPa~\cite{walsh_discovery_2016}.
We note that the discovery of \ce{FeBi2} resulted from extensive MHM structural
searches performed at pressures of 0, 50 and 100~GPa. The most promising
candidate structures were then relaxed at pressure intervals of 10~GPa to
compute enthalpies, which were in turn used to calculate the pressure range of
stability of various phases (see top panel in Fig.~\ref{fig:binaries}a). Besides
the \ce{FeBi2} \textit{I4/mcm} phase, we find a \ce{FeBi3} phase with the
\textit{Cmcm} symmetry to be stable in a very small pressure window slightly
below 40~GPa. This phase has so far not been observed in experiment.


We now compare the pressure range of stability calculated explicitly above
against that calculated using the linear approximation to enthalpy, using only
the ambient equilibrium energies and volumes of the phases. The phase diagram
predicted by the approximate model (bottom panel in Fig.~\ref{fig:binaries}a) is
qualitatively similar to the exact one: the \ce{FeBi2} \textit{I4/mcm} phase
becomes stable at comparable pressures. This finding can be conveniently
exploited in structural searches: since the MHM samples many low-lying
metastable structures at a fixed pressure $p_0$, one could use the energies and
volumes at $p_0$ of such phases within the LAE to quickly predict if any of the
metastable phases become stable at a different pressure $p \ne p_0$. Even for
immiscible systems at $p_0$, potential candidate structures are found if the
simulation cells are sufficiently small to prevent phase segregation. This means
that a structural search conducted solely at 0~GPa might have been sufficient to
uncover the \textit{I4/mcm} structure and correctly predict the experimentally
observed \ce{FeBi2} phase. The \ce{FeBi3} phase, on the other hand, with the
narrow pressure window of stability is predicted to be unstable at all pressures by the
approximate model; this could be a result of errors associated with DFT
calculations. The good agreement between the exact and approximate phase
diagrams is rather surprising: \ce{FeBi2} undergoes a series of magnetic
transitions between 0 and 40~GPa, accompanied by abrupt changes in the unit cell
volume~\cite{amsler_prediction_2017}, all of which are neglected in the linear
approximation to enthalpy.\\

\noindent
\textit{b. Cu--Bi}\\

\noindent
In the ambient-immiscible Cu--Bi system, at least two compounds, with
compositions \ce{Cu11Bi7} and \ce{CuBi}, have been recently discovered in DAC
experiments between 3 and 6~GPa~\cite{clarke_discovery_2016,
clarke_creating_2017}. Both phases can be recovered to ambient conditions, and
exhibit exciting superconducting and structural properties. For example,
\ce{CuBi} has a layered structure, rather uncommon for high-pressure phases, and
is calculated to have an extremely low energy cost associated with exfoliation
from bulk into single sheets~\cite{amsler_cubine_2017}. Further, more recent
structural searches predict additional, dense \ce{Cu2Bi} phases to become
thermodynamically accessible at pressures of above
50~GPa~\cite{amsler_dense_2017}.

The top panel in Fig.~\ref{fig:binaries}b shows the pressure range of stability
of the various high-pressure Cu--Bi phases computed using explicitly calculated
enthalpies for each phase. The \ce{CuBi} phase is not thermodynamically stable
at any pressure at zero temperature, consistent with recent reports of
vibrational entropy playing a crucial role in rendering this phase
synthesizeable~\cite{clarke_creating_2017}. The \ce{Cu11Bi7} phase is
thermodynamically accessible at high pressures up to around 60~GPa, when it
starts to compete with two dense \ce{Cu2Bi} phases~\cite{amsler_dense_2017}.

%

The bottom panel in Fig.~\ref{fig:binaries}b shows the Cu--Bi phase diagram
computed from the LAE, using only the respective equilibrium energy and volume
of each phase at 0~GPa. All phases are correctly predicted to be stable by the
approximate model, consistent with the exact phase diagram. As expected, the
transition pressures predicted by the approximate model are underestimated
overall when compared to those calculated explicitly---a trend  that is
presumably increased due to the significant structural changes in elemental
bismuth as a function of pressure (see Section~\ref{sssec:elemental_solids}).
Nonetheless, it is striking that, using the simple linear approximation to
enthalpy, we could have correctly predicted \textit{all} the high-pressure
phases in the Cu--Bi system from a structural search only at 0~GPa.\\

\noindent
\textit{c. Ni--Bi}\\

\noindent
We tested for the first time the predictive power of our model by investigating
the high-pressure phases in the Ni--Bi binary intermetallic system. Two
compounds have been experimentally reported at ambient pressures: \ce{NiBi} in
the hexagonal NiAs structure~\cite{haegg1929}, and \ce{NiBi3} in the orthogonal
\ce{RhBi3} structure~\cite{glagoleva1954,ruck2006}. Both compounds are
superconductors with transition temperatures of \SI{4.25}{\kelvin} and
\SI{4.06}{\kelvin} in NiBi~\cite{alekseevskii1952} and
\ce{NiBi3}~\cite{alekseevskii1948,herrmannsdorfer2011}, respectively.
To generate phase data to be used within the LAE to construct the convex hull
and predict transition pressures, we used prototypes from our previous
structural searches of the Fe--Bi and Cu--Bi systems, and substituted the Fe/Cu
sites with Ni atoms, followed by structural relaxation at ambient pressures.
Using this ambient-pressure dataset of energies and volumes, the LAE model
predicted stable compounds at high-pressure for the compositions \ce{Ni3Bi} and
\ce{NiBi2}. Based on this prediction, we performed a thorough investigation of
the Ni--Bi system using MHM simulations at pressures of 10 and 50~GPa, which
indeed revealed a number of high-pressure phases.

In particular, our calculations predict new compounds stable at high-pressure at
compositions of the previously reported ambient-pressure phases, i.e., \ce{NiBi}
and \ce{NiBi3}. The hexagonal $\alpha$-NiBi phase undergoes a structural
transition to a TlI-type structure with $Cmcm$ symmetry at pressures above
around 20~GPa. Similarly, the orthorhombic \ce{NiBi3} phase is thermodynamically
unstable above 7.5~GPa, and a $Cmcm$ structure is stable above 62~GPa. Further,
we discover additional stable phases at previously unexplored compositions. We
find that a \ce{NiBi2} phase with $C2/m$ symmetry in the \ce{PdBi2} structure
type is in fact thermodynamically stable at ambient pressures, a finding that
was reported earlier by Bachhuber~\textit{et al.}~\cite{bachhuber_phase_2013}.
At the same composition, a second $C2/m$ phase becomes stable above 52~GPa, over
a very small pressure window of less than 1~GPa, followed by a $I4/mcm$ phase,
isostructural to \ce{FeBi2}. Finally, a \ce{Ni3Bi} compound with $Pmmn$
symmetry, isostructural to \ce{Ni3Sb} in the \ce{Cu3Ti} structure type, is
predicted to be stable at pressures above 25~GPa.

One of our predictions was very recently verified by compressing NiBi in a
diamond anvil cell (DAC). Heating to temperatures above 700$^\circ$C at pressures above $\approx 28$~GPa, the hexagonal
$\alpha$-NiAs transforms into $\beta$-NiAs in the predicted TlI structure
type~\cite{powderly_high-pressure_2017}. The experimental transition pressure is
somewhat higher than the calculated value of 20~GPa. This discrepancy could be
attributed to the presence of high kinetic reaction barriers in the first-order
phase transition, which requires heat to induce the phase change and inevitably leads to calculated transition pressures being
lower than those observed in experiment. This hypothesis is supported by
detectable evidences of the $\beta$-NiAs phase in the XRD pattern upon
decompression: the $\beta$-NiAs is kinetically persistent as low as 11.62~GPa,
hence the equilibrium pressure lies anywhere between 11.62 and 28.3~GPa. In
addition, errors inherent to the approximations used in DFT calculations could
also explain the difference in the observed and computed transition pressures.
The approximations to the exchange correlation potential alone can make a
noticeable difference. E.g., the PBE functional predicts that both the
experimentally observed NiBi and \ce{NiBi3} phases (in their reported
structures) are not thermodynamically stable at 0~GPa and 0~K. However, we find
that LDA correctly places the two experimental phases on the 0~GPa convex hull,
and if we additionally take into account the vibrational entropy contributions
to the free energy, \ce{NiBi2} becomes unstable at elevated temperatures. A
detailed investigation of the influence of different exchange correlation
potentials and temperature effects on the calculated phase stability of Ni--Bi
compounds and their properties will be reported elsewhere.


After exploring the high-pressure Ni--Bi system with the MHM, we \textit{a
posteriori} compare the phase diagram of Ni--Bi computed using the explicitly
calculated enthalpies against that predicted from our LAE model
(Fig.~\ref{fig:binaries}c), and find remarkable agreement. Most phases, and the
sequence in which they form under pressure, are correctly predicted by the
approximate model. The only exceptions are the $Cmcm$ phase at the \ce{NiBi3}
composition and the second $C2/m$ compound at the \ce{NiBi2} composition at
around 50~GPa. As discussed earlier, the latter phase has a very small pressure
range of stability of $<$1~GPa, so its absence in the phase diagram predicted by
the approximate model is not surprising. In fact, similar to the $Cmcm$
\ce{FeBi3} phase that was predicted to be stable in a narrow pressure window of
less than 3~GPa but not yet observed experimentally, synthesis of the \ce{NiBi2}
phase is likely to be challenging, if possible at all.

%% file: statistics_elements.tex
The power of our linear enthalpy model lies in its capability to efficiently
assess the pressure range of stability of hundreds of thousands of phases. Since
the linear approximation requires only equilibrium energies and volumes of
phases calculated at ambient pressure, it can be used to leverage the large
materials datasets available in HT-DFT databases such as the
OQMD~\cite{saal_materials_2013, kirklin_oqmd_2015}, Materials
Project~\cite{Jain2013}, and AFLOWlib~\cite{curtarolo_aflowlib.org_2012}. Here,
we present large-scale analysis and statistics of thermodynamic phase
stability of materials at high pressure using ambient-pressure phase data
calculated in the OQMD.

First, we focus on elemental high-pressure phases, and begin by compiling a
``validation-dataset'' of experimentally reported high-pressure elemental phases.
The crystal structures of many high-pressure phases reported in the Inorganic
Crystal Structure Database (ICSD)~\cite{icsd} have been calculated in the OQMD,
albeit at ambient pressure.  For every element, we filter all entries in the
ICSD using the ``External Conditions $\rightarrow$ Pressure'' metadata available
for each entry. Further, Tonkov~et~al.~\cite{tonkov_phase_2004} compiled a
comprehensive list of phase transformations under pressure for nearly 100
elements, on which we rely heavily as a second reference to cross-validate and
augment the list of high-pressure phases calculated in the OQMD. Our final
compiled dataset contains 132 distinct elemental high-pressure phases, and can
be found in the Supplementary Materials (SM).

\begin{figure}[!htbp]
\includegraphics[width=0.9\columnwidth]{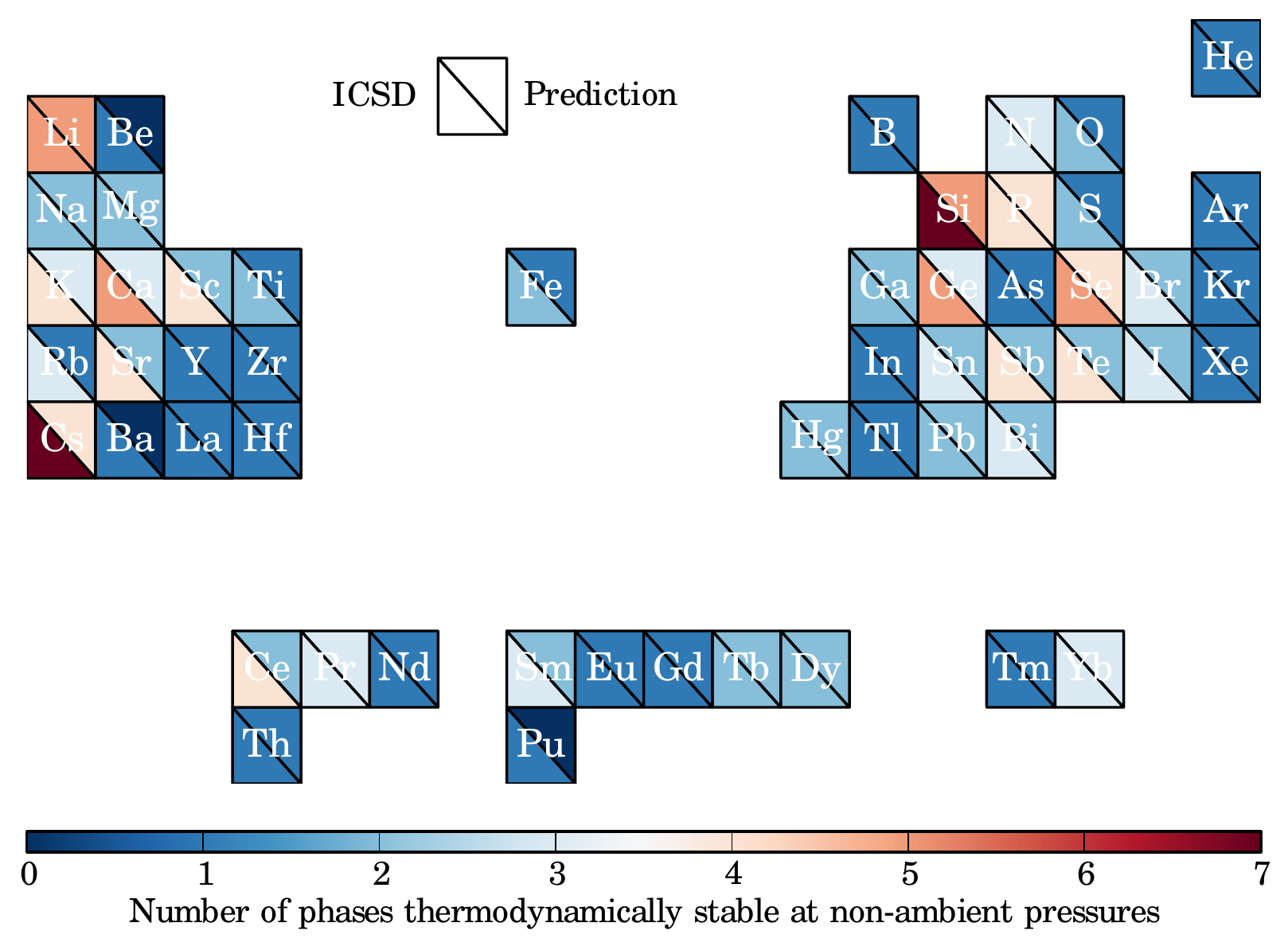}
\caption{Comparison of predictions of high-pressure elemental phases from the
LAE model against experiment. For each element, the number of (a)
unique phases reported experimentally and (b) predicted by the linear enthalpy
model to be thermodynamically stable at non-ambient pressures, are indicated by
the color of the bottom-left and top-right halves, respectively, of the
corresponding tile in the periodic table. Overall, the model correctly predicts
$\sim$75\% of the high-pressure phases in the ICSD to be thermodynamically
stable at non-ambient pressures.}
\label{fig:elements_ICSD_vs_prediction}
\end{figure}

For each element in the periodic table, we use the ambient-pressure energy and
volume data for \textit{all} ICSD phases (i.e., not limited to high-pressure
phases) calculated in the OQMD within the LAE model to predict (a) the number of
phases from our validation-dataset that lie on the extended $N$--$V$--$E$ convex
hull, i.e., the number of phases stable at some finite pressure, and (b) the
pressure range of stability of every phase that lies on the $N$--$V$--$E$ hull.
Fig.~\ref{fig:elements_ICSD_vs_prediction} shows a summary of this analysis in
the form of a periodic table: for every element with at least one experimentally
reported high-pressure phase, we indicate the number of high-pressure phases in
our compiled dataset from the OQMD (bottom-left half) and the number of phases
predicted by the linear enthalpy model to lie on the $N$--$V$--$E$ hull
(top-right half), represented on a color scale. That is, the number of phases
reported experimentally and those predicted to be stable at some finite-pressure
match exactly whenever the colors in both the left and right segments are
identical. This is indeed the case for most elements, with a few exceptions.
Overall, 75\% of all experimentally reported high-pressure phases are predicted
to lie on the $N$--$V$--$E$ convex hull (see the top panel of
Table~\ref{tab:statistics}). In addition, for around 35\% of the phases, the
predicted pressure range of stability overlaps with the respective transition
pressures reported in experiment. The low success rate in correctly predicting
the transition pressure is somewhat expected following the model validation on
Si and Bi in Section~\ref{ssec:model_validation}. We discuss the possible
sources of discrepancy between predictions and experiment toward the end of this
Section.

Next, we perform a similar large-scale analysis for all experimentally reported
binary phases. Using calculations of experimentally reported compounds in the
OQMD, curated using pressure-related metadata in the ICSD (in a manner similar
to that employed for elemental phases), we compile a dataset of 343 unique
binary compounds in total as a validation-dataset (the entire list is available
in Supplementary Materials). This number is significantly lower than that
expected from a simple combinatorial estimation. For elemental solids, we found
in average more than one high-pressure phase per element. If we extend this
observation to binaries and assume that every binary system has in average more
than one high-pressure phase, the number of potential high-pressure phases
considering 90 elements is  $^{90}{\rm C}_2 = 4005$. We note that our estimation
is very conservative, since binary $A$--$B$ systems introduce an additional,
compositional degree of freedom, which allows multiple high-pressure phases to
exist at the same pressure, \ce{A_pB_q}, as we have seen in
Sec.~\ref{sssec:binary_intermetallics}. This indicates that the high-pressure
phase diagrams of binary systems in general have been relatively underexplored.
The linear enthalpy model performs equally well for binary compounds --- 80\% of
experimentally reported high-pressure binary phases are predicted to be stable
at some finite pressure (see lower panel of Table~\ref{tab:statistics}). For
around 35\% of the phases, the predicted pressure range of stability overlaps
with the respective transition pressures reported experimentally.

\begin{table}
\caption{Accuracy of the linear enthalpy model in predicting the stability (at
some finite-pressure) of experimentally reported elemental and binary
high-pressure phases.\label{tab:statistics}}
\begin{tabular}{lr}
\toprule
\multicolumn{2}{c}{Elements} \\
\midrule
Experimentally reported HP phases & 132 \\
Predicted to be stable at finite pressure & 97 (75\%) \\
Predicted pressure range of stability & \multirow{2}{*}{45 (35\%)} \\
matches experiment & \\
\midrule
\multicolumn{2}{c}{Binaries} \\
\midrule
Experimentally reported HP phases & 343 \\
Predicted to be stable at finite pressure & 273 (80\%) \\
Predicted pressure range of stability & \multirow{2}{*}{125 (35\%)} \\
matches experiment & \\
\bottomrule
\end{tabular}
\end{table}

Overall, our ``crude'' linear enthalpy model performs surprisingly well, with a
success rate of 75--80\%, in predicting the stability of both elemental and
binary high-pressure phases. We identify four potential sources of error that
could explain the discrepancy between the number of high-pressure phases
reported experimentally and that predicted by our approximate model:
\begin{enumerate}
\item [(a)] The crystal structure reported experimentally for the phase is
erroneous. Resolving the crystal structure, e.g., from in-situ XRD measurements,
under high pressure is a difficult and tedious task that can lead to
incomplete/incorrect structural characterization. A prominent example is the
Bi-III phase, the crystal structure of which was experimentally resolved only
after several failed attempts~\cite{chen_crystal_1994}. In fact, Bi-III has an
incommensurate host-guest structure and the reported structure is only a
representative ordered model with $P4/nnc$
symmetry~\cite{haussermann_comparative_2002}. A similar incommensurate structure
has been reported for phase IV of phosphorus in the pressure range of
107-137~GPa~\cite{fujihisa_incommensurate_2007}.
\item [(b)] The high-pressure phase emerges via a phase transition of second
order. In this case, the structural relaxations performed using DFT will
inevitably transform the high-pressure phase to a lower-pressure structure.
Therefore, our linear enthalpy model, which relies on the equilibrium energy
$E_0$ and volume $V_0$ at ambient pressure of a high-pressure phase, will
expectedly not capture its stability.
\item [(c)] Errors inherent to DFT calculations and numerical noise, e.g., due
to the approximation to the exchange correlation potential, pseudization of core
electrons (which might be important especially at high pressures), unconverged
basis sets and sampling meshes, insufficient tolerances during structural
relaxations, etc.
\item [(d)] Finally, there is the inherent error due to applying a linear
approximation to the enthalpy of each phase (i.e., assuming all phases to be
perfectly incompressible), which might be unreasonable for some materials at
large values of pressure.
\end{enumerate}

%% file: statistics_all.tex
We now use our linear enthalpy model to analyze the phase stability of
all experimentally reported compounds calculated in the OQMD (not
limited to high-pressure phases), a total of around 33,000 unique ordered
compounds. As earlier, using the equilibrium energy and volume at ambient
pressure of each phase in our dataset, we predict the number, and the pressure
range of stability, of all phases that lie on the extended $N$--$V$--$E$ convex
hull (i.e., presumably thermodynamically stable at some finite-pressure).

First, we find that only around 55\% of the 33,000 compounds in our dataset lie
on the $N$--$E$ convex hull, i.e., are thermodynamically stable at ambient
pressure conditions, consistent with a previous report on a similar dataset from
the OQMD~\cite{kirklin_oqmd_2015}. A recent study by
Sun~et~al.\cite{sun_thermodynamic_2016} on a dataset of 29,900 experimentally
reported compounds calculated in the Materials Project also finds around
50$\pm$4\% of the phases to be ambient-metastable. In the latter study, it is
proposed that the \textit{observed} metastable compounds are generally remnants
of thermodynamic conditions where they were once the stable ground states.

\begin{figure}[!htbp]
\includegraphics[width=0.9\columnwidth]{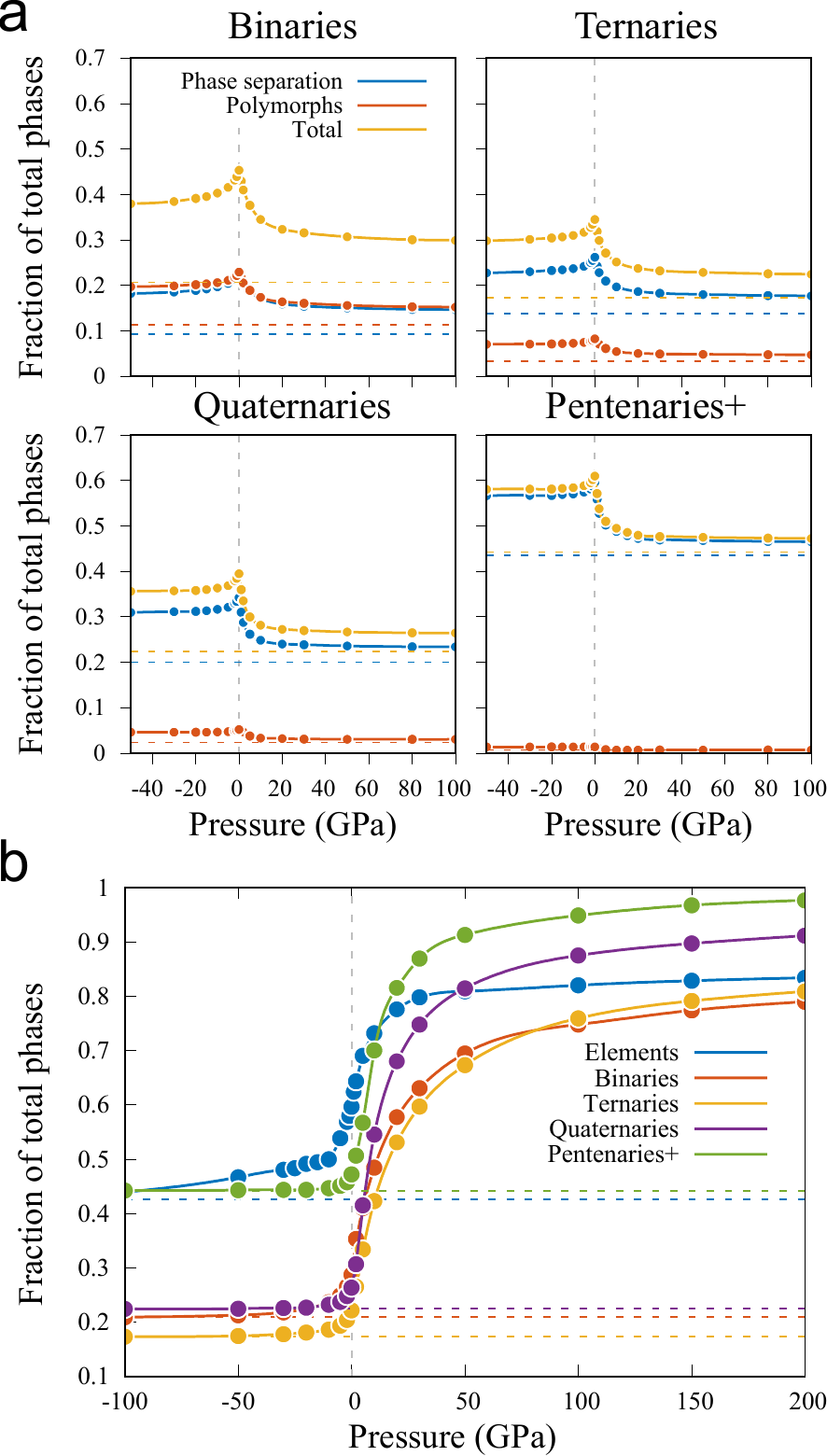}
\caption{(a) Fraction of metastable phases that become thermodynamically stable
with incremental increase/decrease in pressure, with respect to 0~GPa. The
horizontal dashed lines indicate the fraction of metastable phases that do not
lie on the $N$--$V$--$E$ convex hull at any pressure. (b) Fraction of
ambient-metastable phases that cannot be accessed thermodynamically at any
pressure larger than pressure $p$, equivalent to $1-$ (fraction of phases that
can be accessed at some pressure larger than pressure
$p$).}\label{fig:ICSD_stat}
\end{figure}

We next test this hypothesis of ``remnant metastability'' by using pressure as a
thermodynamic handle and tracking the number of metastable phases that become
stable with incremental increase/decrease in pressure, with respect to ambient
conditions. Fig.~\ref{fig:ICSD_stat}a shows the fraction of metastable
phases as a function of positive (compressive) or negative (tensile) pressure,
separated into binary, ternary, quaternary and higher-component systems. We
observe a range of trends based on our statistical analysis.

First of all, the number of metastable phases decreases with incremental
application of both positive and negative pressures, relative to 0~GPa. In other
words, \textit{a significant fraction of the ambient-metastable phases are in
fact thermodynamically stable ground states at non-ambient pressure conditions.}
For example, in the case of binary compounds (top left in
Fig.~\ref{fig:ICSD_stat}a), the fraction of metastable phases decreases from
around 0.45 at 0~GPa to around 0.30 at 100~GPa---15\% of the ambient-metastable
phases are rendered thermodynamically stable at some pressure $p \in (0,
100]$~GPa. However, in each case, a sizeable fraction of metastable phases
remain metastable at all pressures, i.e., they are not equilibrium ground states
at any pressure (represented by horizontal dashed lines in
Fig.~\ref{fig:ICSD_stat}a). For example, around 21\% of all binary
ambient-metastable phases cannot be accessed thermodynamically via pressure
alone.

Second, the rate of decrease in the number of metastable phases (or increase in
the number of metastable phases made stable) with pressure is maximum near zero
and decays rapidly toward higher positive/negative pressures. This is most
likely due to a bias toward small values of pressure in our compiled set of
phases---after all, most compounds reported experimentally are likely observed
in near-ambient conditions---but could be also due to a fundamental property of
materials, namely, the density of stable ground states as a function of
volume/pressure is maximum near zero pressure.

Third, we find considerable differences concerning the ``character'' of
metastability in binary, ternary, and higher order compositional systems. We
distinguish two subsets for each $n$-component dataset ($n=$ 2, 3, 4,
$\ge$5)---``polymorphs'' and ``phase separation''---depending on whether a given
phase is metastable with respect to another phase at the same composition or a
combination of phases, respectively, at ambient conditions. We note that the
higher the number of components present in a metastable compound, the more
likely it is to phase-separate rather than transform into a polymorph, in
agreement with previous observations~\cite{sun_thermodynamic_2016}. Further, the
lower the number of components in a metastable compound, the more likely it is
to be stabilized with pressure. Considering the subset of all metastable phases
that phase-separate at ambient pressure, 58\%, 47\%, 42\%, and 27\% become
thermodynamically stable at some finite positive/negative pressure in the case
of binary, ternary, quaternary, and higher-component systems, respectively.

Additionally, we observe that the effects of positive and negative pressures on
the metastability of phases are \textit{not} symmetric about zero pressure: a
much larger portion of ambient-metastable phases become thermodynamically stable
under positive (compressive) pressure when compared to negative (tensile)
pressure. A difference is perhaps expected considering that the limiting
behaviors are very different: large positive pressures favor the formation of
close-packed phases before eventual overlap of atomic cores, while the limit of
large negative pressures is simply the individual non-interacting atoms of each
species in the phase.


Finally, we probe a complementary question: if one were to incrementally tune
external conditions from large positive to large negative pressures, how many
observed metastable phases $\mathcal{N}$ can be accessed thermodynamically below
any given pressure $p$? We calculate at pressure $p$, the number of
experimentally observed phases from our dataset that \textit{cannot} be
thermodynamically accessed at any pressure $>p$. We present this data as a
cumulative histogram of the fraction of phases, integrated from pressures $p$ to
$+\infty$, separated into elements, binaries, ternaries, etc.\ in
Fig.~\ref{fig:ICSD_stat}b. Hypothetically, if all experimentally reported
compounds were thermodynamically stable ground states at some finite pressure,
one would expect this cumulative fraction of unstable phases to be 1 and 0 for
$p \to \infty$ and $p \to -\infty$, respectively. Consistent with our previous
observations, we find that (a) a sizable fraction of the phases do not lie on
the extended $N$--$V$--$E$ convex hull at all, i.e., they are not ground states
under any pressure (represented by horizontal dashed lines in
Fig.~\ref{fig:ICSD_stat}b), and (b) the rate at which additional metastable
phases can be thermodynamically accessed is maximum near zero pressure (given by
the slopes of the curves). In other words, the pressure density of thermodynamic
ground states, $\frac{d\mathcal{N}}{dp}$, is maximum near $p = 0$. Whether this
is an artifact of using a dataset of experimentally observed phases or is a
fundamental property of matter, needs further analysis, and will be the subject
of future work.


%% file: binary_immiscible.tex
So far, we have used the LAE to analyze the phase stability of experimentally
reported high-pressure elemental and binary phases, and to probe the
accessibility of ambient-metastable phases using pressure as a thermodynamic
handle. Now, we go a step further by using the LAE to predict new intermetallic
compounds by combining it with CSP methods. For this purpose, we focus on a
unique subset of binary systems, namely, the combination of elements that are
immiscible at ambient pressures. According to the data we compiled from the
OQMD, there currently exist $\sim$1780 binary systems that do not contain any
experimentally observed compounds. Any high-pressure phases that we identify in
these systems are therefore true predictions of new materials.

\begin{figure}[!htbp]
\begin{center}
\includegraphics[width=1.\columnwidth]{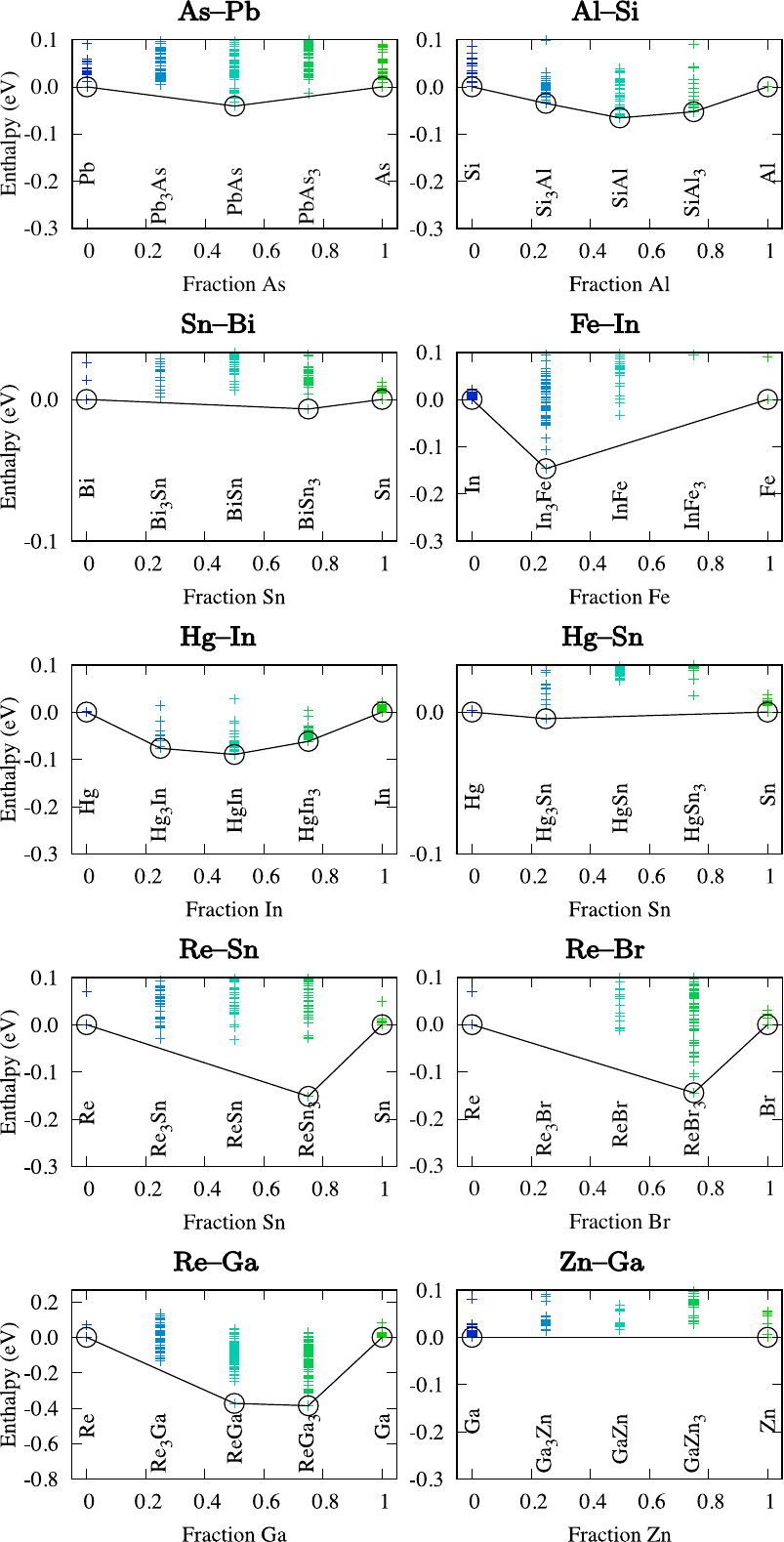}\label{fig:testset}
\end{center}
\caption{The convex hulls of formation enthalpy of ten ambient-immiscible binary systems calculated using structural
search at 50~GPa via the MHM. Each cross denotes a phase sampled with the MHM. In all but the Zn-Ga system, we find at
least one thermodynamically stable high-pressure phase.}\label{fig:binary_predictions}
\end{figure}

For the dataset to be used for construction of the convex hull and calculation of
transition pressures within the LAE, we use ambient-pressure formation energies
and volumes of phases calculated in the OQMD. As mentioned in
Section~\ref{ssub:calculation_settings}, the OQMD contains calculations of more
than 450,000 compounds including experimentally reported compounds from the
ICSD, and hypothetical compounds generated by decoration of common structural
prototypes with all the elements in the periodic table. The Strukturbericht
symbols of the prototype structures considered in this section are listed
below~\cite{kirklin_oqmd_2015, kirklin_high-throughput_2016}:
\begin{enumerate}
\item[(a)] elemental prototypes: A1 (fcc), A2 (bcc), A3 (hcp), A3'
($\alpha$-La), A4 (diamond), A5 ($\beta$-Sn), A7 ($\alpha$-As), A9 (graphite),
A10 ($\alpha$-Hg), A11 ($\alpha$-Ga), A12 ($\alpha$-Mn), A13 ($\beta$-Mn),
A$_{\rm b}$ ($\beta$-U), A$_{\rm h}$ ($\alpha$-Po), $C19$ ($\alpha$-Sm)
\item[(b)] binary $AB$: B1 (NaCl), B2 (CsCl), B3 (zincblende ZnS), B4 (wurtzite
ZnS), B19 (AuCd), B$_h$ (WC), L1$_0$ (AuCu), L1$_1$ (CuPt) 
\item[(c)] binary $A_3B$: L1$_2$ (Cu$_3$Au), D0$_{19}$ (Ni$_3$Sn), D0$_{22}$
(Al$_3$Ti), D0$_3$ (AlFe$_3$)
\end{enumerate}

We screen for promising chemical systems that contain high-pressure phases in
the following manner: for every ambient-immiscible binary system, we use the LAE
to predict the thermodynamic phase stability and pressure range of stability of
each hypothetical compound in that chemical space. We select systems that
contain at least one hypothetical compound predicted to become stable below an
arbitrary pressure threshold of 50~GPa. We then rank these systems according to
the predicted transition pressures, from lowest to highest, and select 10 of the
most promising systems for further investigation. For each system, we further
verify that no compound in that chemical space is reported in the ICSD or in
phase diagrams available in the ASM Alloy Phase Diagram
Database~\cite{massalski_binary_1986}. At each composition where our model
predicts a stable high-pressure phase we perform structural searches using the
MHM, starting from the respective prototype structure from the OQMD, using
simulation cells with up to 10 atoms/cell. Due to the set of binary prototypes
currently calculated in the OQMD (see list above), the compositions we sample
are limited to $A_3B$, AB and $AB_3$. Note that both the system size and the
number of sampled compositions are far too low to give accurate predictions of
the true high-pressure ground states. The structural searches are merely
intended as proof-of-concept, i.e., to provide a sampling of configurations
beyond the limited number of prototype structures.

Of the ten selected ambient-immiscible binary systems, namely, As-Pb, Al-Si,
Sn-Bi, Fe-In, Hg-In, Hg-Sn, Re-Sn, Re-Br, Re-Ga, and Zn-Ga, structural searches
performed at 50~GPa using the MHM confirmed the existence of at least one new
stable high-pressure phase in all but the Zn-Ga system (see
Fig.~\ref{fig:binary_predictions}). All thermodynamically stable structures at
50~GPa are provided in the Supplementary Materials (SM). The high-pressure
phases predicted present a number of avenues for experimental synthesis and
verification. Overall, the success of the linear enthalpy model in guiding more
accurate, sophisticated techniques based on crystal structure prediction in
discovering novel high-pressure phases is remarkable.

%% file: conclusions.tex
In summary, we present a method that allows an efficient screening for materials
that are thermodynamically stable at non-ambient pressures using a simple linear
approximation to the formation enthalpy of a phase. Using a generalized convex hull construction, the
stability of thousands of compound can be evaluated at a low computational cost based on
ambient-pressure data that is currently available in many materials databases
without performing any additional DFT calculations. Through a large-scale
analysis of experimentally reported compounds, we show that a large fraction of
the observed ambient-metastable phases are in fact thermodynamic ground states
at some finite pressure. Our method can be readily extended by further generalizing the convex hull construction and taking into account additional thermodynamic degrees of freedom, including temperature or surface areas of finite particles. Finally, we demonstrate the predictive power of this
model when combined with a crystal structure prediction technique by discovering
novel high-pressure phases in a set of ambient-immiscible binary intermetallic
systems. 

%% file: method2.tex
\subsection{Calculation of thermodynamic quantities}\label{ssub:calculation_settings}

The equilibrium formation energy and volume data for all the phases considered
in our analysis using LAE were retrieved from the Open Quantum Materials Database
(OQMD)~\cite{saal_materials_2013, kirklin_oqmd_2015}. The dataset consists of
DFT-calculated properties of over 450,000 compounds which include (a) unique,
ordered experimentally reported compounds from the Inorganic Crystal Structure
Database (ICSD), and (b) hypothetical compounds generated by the decoration of
common structural prototypes with all the elements in the periodic table.
Details of the settings used to calculate the equilibrium formation energy and
volume of compounds in the OQMD can be found in
Ref.~\onlinecite{kirklin_oqmd_2015}.

All other DFT calculations reported in this work, i.e., those performed as part
of global structure searches, were performed using the Vienna Ab initio
Simulation Package (VASP)~\cite{kresse1993ab, kresse1996efficiency,
kresse1996efficient}. We use the projector augmented wave (PAW)
formalism~\cite{blochl1994projector, kresse_paw_1999} and the PBE
parameterization of the generalized gradient approximation to the exchange
correlation functional~\cite{perdew1996generalized} throughout. For all
calculations, we use $\Gamma$-centered $k$-point meshes with about 8000
$k$-points per reciprocal atom and a plane-wave cutoff energy of 520~eV. All
atomic and cell degrees of freedom of a structure are relaxed until the force
components on all the atoms are within 0.01~eV/\AA{}, and stresses are within a
few kbar.

\subsection{Structural searches}\label{ssub:structure_search_mhm}

The minima hopping method (MHM)~\cite{goedecker_minima_2004,
amsler_crystal_2010} implements a highly reliable algorithm to explore the low
enthalpy phases of a compound at a specific pressure given solely the chemical
composition~\cite{amsler_novel_2012, huan_low-energy_2012,
huan_thermodynamic_2013}. The low lying part of the enthalpy landscape is
efficiently sampled by performing consecutive, short MD escape steps to overcome
enthalpy barriers, followed by local geometry optimizations. The
Bell-Evans-Polanyi principle is exploited through a feedback mechanism on the MD
escape trials, and by aligning the initial MD velocities along soft-mode
directions in order to accelerate the search~\cite{roy_bell-evans-polanyi_2008,
sicher_efficient_2011}. The MHM has been successfully applied to identify the
structure and composition of many materials, also for systems at high
pressures~\cite{amsler_crystal_2012, flores-livas_high-pressure_2012,
flores_Superconductivity_2016, amsler_prediction_2017, clarke_creating_2017,
amsler_dense_2017}. In this work, we performed MHM simulations only at the
compositions where a high-pressure phase is predicted to be stable by the linear
enthalpy model.

\subsection{Software implementation}\label{ssub:software}

All convex hull constructions in this work were performed using the
\texttt{Qhull} library~\cite{barber96} as implemented in the \texttt{SciPy}
Python package~\cite{jones2001}.  The GCLP calculations reported in this work
were performed using the \texttt{Cbc} solver distributed with the \texttt{PuLP}
Python library~\cite{Mitchell11}. An implementation of the framework described
in
Sections~\ref{ssec:linear_approximation_to_enthalpy}--\ref{ssec:pressure_range_of_stability}
has been made available as an open-source Python module~\cite{Hegde2017}. An
implementation of the MHM is available through the \texttt{Minhocao}
package~\cite{goedecker_minima_2004, amsler_crystal_2010}.

%% file: acknowledgements.tex
M.A.\ and V.I.H. conceived and carried out the project, and contributed equally
to this work, S.D.J.\ and C.W.\ supervised the project, and all authors
contributed to writing the manuscript. M.A.\ (construction of linear model,
crystal structure prediction) acknowledges support from the Novartis
Universit{\"a}t Basel Excellence Scholarship for Life Sciences and the Swiss
National Science Foundation (Project No.\ P300P2-158407, P300P2-174475).
V.I.H.\ (model implementation, high-throughput calculations) and C.W.\
acknowledge support from the Department of Energy, Office of Science, Basic
Energy Sciences under Grant DE-SC0015106. 
S.D.J. acknowledges support from NSF DMR-1508577 and the
Carnegie/DOE Alliance Center (CDAC).
The authors acknowledge support from the Data Science Initiative at
Northwestern University. The computational resources from the Swiss National
Supercomputing Center in Lugano (Projects s499, s621, s700), the Extreme Science
and Engineering Discovery Environment (XSEDE) (which is supported by National
Science Foundation Grant OCI-1053575), the Bridges system at the Pittsburgh
Supercomputing Center (PSC) (which is supported by NSF Award ACI-1445606), the
Quest high performance computing facility at Northwestern University, and the
National Energy Research Scientific Computing Center (DOE: DE-AC02-05CH11231),
are gratefully acknowledged.